\documentclass[aps,prc,twocolumn,showpacs,superscriptaddress,floatfix,10pt]{revtex4-1}
\usepackage{graphicx}
\usepackage{dcolumn}
\usepackage{bm}
\usepackage{amsfonts}
\usepackage{amssymb}
\usepackage{amsmath}

\begin{document}

\title{Microscopic analysis of sub-barrier fusion enhancement in $^{132}$Sn+$^{40}$Ca vs. $^{132}$Sn+$^{48}$Ca}

\author{V.E. Oberacker}
\author{A.S. Umar}
\affiliation{Department of Physics and Astronomy, Vanderbilt University, Nashville, Tennessee 37235, USA}

\date{\today}


\begin{abstract}
We provide a theoretical analysis of recently measured fusion cross sections
which show a surprising enhancement at low $E_\mathrm{c.m.}$ energies
for the system $^{132}$Sn+$^{40}$Ca as compared to the more neutron-rich system
$^{132}$Sn+$^{48}$Ca. Dynamic microscopic calculations are carried out on a
three-dimensional lattice with a time-dependent density-constrained density
functional theory. There are no adjustable parameters, the only input is the
Skyrme effective NN interaction. Heavy-ion potentials $V(R)$, coordinate-dependent mass parameters $M(R)$,
and total fusion cross sections $\sigma(E_\mathrm{c.m.})$ are calculated for both systems.
We are able to explain the measured fusion enhancement in terms of the {\it narrower width}
of the ion-ion potential for $^{132}$Sn+$^{40}$Ca, while the barrier heights and positions
are approximately the same in both systems.
\end{abstract}
\pacs{21.60.-n,21.60.Jz}
\maketitle


\section{Introduction}

Radioactive ion beam facilities enable us to study fusion reactions with
exotic neutron-rich nuclei. An important goal of these experiments is to
study the effects of neutron excess ($N-Z$) on fusion.
In several experiments, large enhancements of sub-barrier fusion yields have
been observed for systems with positive Q values for neutron transfer.
Recently, at the HRIBF facility a series of experiments has been carried out
with radioactive $^{132}$Sn beams and with stable $^{124}$Sn beams on
$^{40,48}$Ca targets~\cite{KR12}. It turns out that the $^{40}$Ca+Sn systems
have many positive Q values for neutron-pickup while all the Q values for
$^{48}$Ca+Sn are negative. However, the data analysis reveals that the fusion
enhancement is not proportional to the magnitudes of those Q values.

Particularly puzzling is the experimental observation of a sub-barrier fusion
enhancement in the system $^{132}$Sn+$^{40}$Ca
as compared to more neutron-rich system $^{132}$Sn+$^{48}$Ca.
This is difficult to understand because the $8$ additional neutrons in
$^{48}$Ca should increase the attractive strong nuclear interaction
and thus lower the fusion barrier, resulting in an enhanced sub-barrier
fusion cross section. But the opposite is found experimentally.
A coupled channel analysis~\cite{KR12} of the fusion data with
phenomenological heavy-ion potentials yields cross sections that are one order
of magnitude too small at sub-barrier energies, despite the fact that
these ion-ion potentials contain many adjustable parameters.

The time-dependent Hartree-Fock (TDHF) theory provides a useful foundation for a
fully microscopic many-body theory of large amplitude collective
motion~\cite{Ne82,Si11} including deep-inelastic and fusion reactions.
But only in recent years has it become feasible to perform TDHF calculations on a
3D Cartesian grid without any symmetry restrictions
and with much more accurate numerical methods~\cite{Si11,US91,UO06,Si12,GM08,DD-TDHF}.
In addition, the quality of effective interactions has been substantially
improved~\cite{CB98,Klu09a,KL10,USR}.
During the past several years, we have developed the Density Constrained
Time-Dependent Hartree-Fock (DC-TDHF) method for calculating
heavy-ion potentials~\cite{UO06a}, and we have applied this method
to calculate fusion and capture cross sections above and below the barrier. As of to date,
we have studied a total of $18$ different systems, including $^{132,124}$Sn+$^{96}$Zr~\cite{OU10b},
$^{132}$Sn+$^{64}$Ni~\cite{UO06d,UO07a},
$^{16}$O+$^{208}$Pb~\cite{UO09b},  and hot and cold fusion reactions leading
to superheavy element $Z=112$~\cite{UO10a}. Most recently, we have investigated
sub-barrier fusion and pre-equilibrium giant resonance excitation between
various calcium + calcium isotopes~\cite{KU12}, and between
isotopes of oxygen and carbon~\cite{UO12} that occur in the neutron
star crust. In all cases, we have found
good agreement between the measured fusion cross sections and the DC-TDHF results.
This is rather remarkable given the fact that the only input in DC-TDHF is the
Skyrme effective N-N interaction, and there are no adjustable parameters.

This paper is organized as follows: in Section~II we summarize the theoretical
formalism and show results for the ion-ion potentials calculated with the
DC-TDHF method. In Section~III, we discuss the corresponding total fusion cross
sections, and we explain the observed fusion enhancement in terms of the
{\it narrower width} of the ion-ion potential for $^{132}$Sn+$^{40}$Ca.
Our conclusions are presented in Section~IV.


\section{Theoretical formalism and ion-ion potentials}

Currently, a true quantum many-body theory of barrier tunneling does not exist.
All sub-barrier fusion calculations assume that there exists an ion-ion potential
$V(R)$ which depends on the internuclear distance $R$. Most of the theoretical
fusion studies are carried out with the coupled-channels (CC) method~\cite{HW07,Das07,IH09,EJ10}
in which one uses empirical ion-ion potentials (typically Woods-Saxon potentials,
or double-folding potentials with frozen nuclear densities).

While phenomenological methods provide a useful starting point for the analysis
of fusion data, it is desirable to use a quantum many-body approach which
properly describes the underlying nuclear shell structure of the reaction system.
We have developed a microscopic approach to extract heavy-ion interaction potentials
$V(R)$ from the TDHF time-evolution of the dinuclear system.
The interaction potentials calculated with the DC-TDHF method incorporate
all of the dynamical entrance channel effects such as neck formation,
particle exchange, internal excitations, and deformation effects~\cite{UO08a}.
While the outer part of the potential barrier is largely determined by
the entrance channel properties, the inner part of the potential barrier
is strongly sensitive to dynamical effects such as particle transfer and neck
formation.

The TDHF equations for the single-particle wave functions
\begin{equation}
h(\{\phi_{\mu}\}) \ \phi_{\lambda} (r,t) = i \hbar \frac{\partial}{\partial t} \phi_{\lambda} (r,t)
            \ \ \ \ (\lambda = 1,...,A) \ ,
\label{eq:TDHF}
\end{equation}
can be derived from a variational principle.  The main approximation in TDHF is
that the many-body wave function $\Phi(t)$  is assumed to be a single time-dependent
Slater determinant which consists of an anti-symmetrized product of single-particle
wave functions
\begin{equation}
\Phi(r_1,...,r_A;t) = (A!)^{-1/2} \ det |\phi_{\lambda} (r_i,t) | \ .
\label{eq:Slater}
\end{equation}
In the present TDHF calculations we use the Skyrme SLy4 interaction~\cite{CB98} for the nucleons
including all of the time-odd terms in the mean-field Hamiltonian~\cite{UO06}.
The numerical calculations are carried out on a 3D Cartesian lattice. For
$^{40,48}$Ca+$^{132}$Sn the lattice spans $50$~fm along the collision axis and $30-42$~fm in
the other two directions, depending on the impact parameter. First we generate very accurate
static HF wave functions for the two nuclei on the 3D grid. In the second
step, we apply a boost operator to the single-particle wave functions. The time-propagation
is carried out using a Taylor series expansion (up to orders $10-12$) of the unitary mean-field propagator,
with a time step $\Delta t = 0.4$~fm/c.

In our DC-TDHF approach, the time-evolution takes place with no restrictions.
At certain times $t$ or, equivalently, at certain internuclear distances
$R(t)$ the instantaneous TDHF density
\begin{equation}
\rho_{\mathrm{TDHF}}(r,t) = <\Phi(t) |\rho| \Phi(t) >
\label{eq:rho_TDHF}
\end{equation}
is used to perform a static Hartree-Fock energy minimization
\begin{equation}
\delta <\Phi_{\rho} \ | H - \int d^3r \ \lambda(r) \ \rho(r) \ | \Phi_{\rho} > = 0
\label{eq:var_dens}
\end{equation}
while constraining the proton and neutron densities to be equal to the instantaneous
TDHF densities
\begin{equation}
<\Phi_{\rho} |\rho| \Phi_{\rho} > = \rho_{\mathrm{TDHF}}(r,t) \ .
\label{eq:dens_constr}
\end{equation}
These equations determine the state vector $\Phi_{\rho}$. This means we
allow the single-particle wave functions to rearrange themselves in such a way
that the total energy is minimized, subject to the TDHF density constraint.

In a typical DC-TDHF run, we utilize a few
thousand time steps, and the density constraint is applied every $10-20$ time steps.
We refer to the minimized energy as the ``density constrained energy'' $E_{\mathrm{DC}}(R)$
\begin{equation}
E_{\mathrm{DC}}(R) = <\Phi_{\rho} | H | \Phi_{\rho} > \ .
\label{eq:EDC}
\end{equation}
The ion-ion interaction potential $V(R)$ is essentially the same as $E_{\mathrm{DC}}(R)$,
except that it is renormalized by subtracting the constant binding energies
$E_{\mathrm{A_{1}}}$ and $E_{\mathrm{A_{2}}}$ of the two individual nuclei
\begin{equation}
V(R)=E_{\mathrm{DC}}(R)-E_{\mathrm{A_{1}}}-E_{\mathrm{A_{2}}}\ .
\label{eq:vr}
\end{equation}

\begin{figure}[!htb]
\includegraphics*[width=8.6cm]{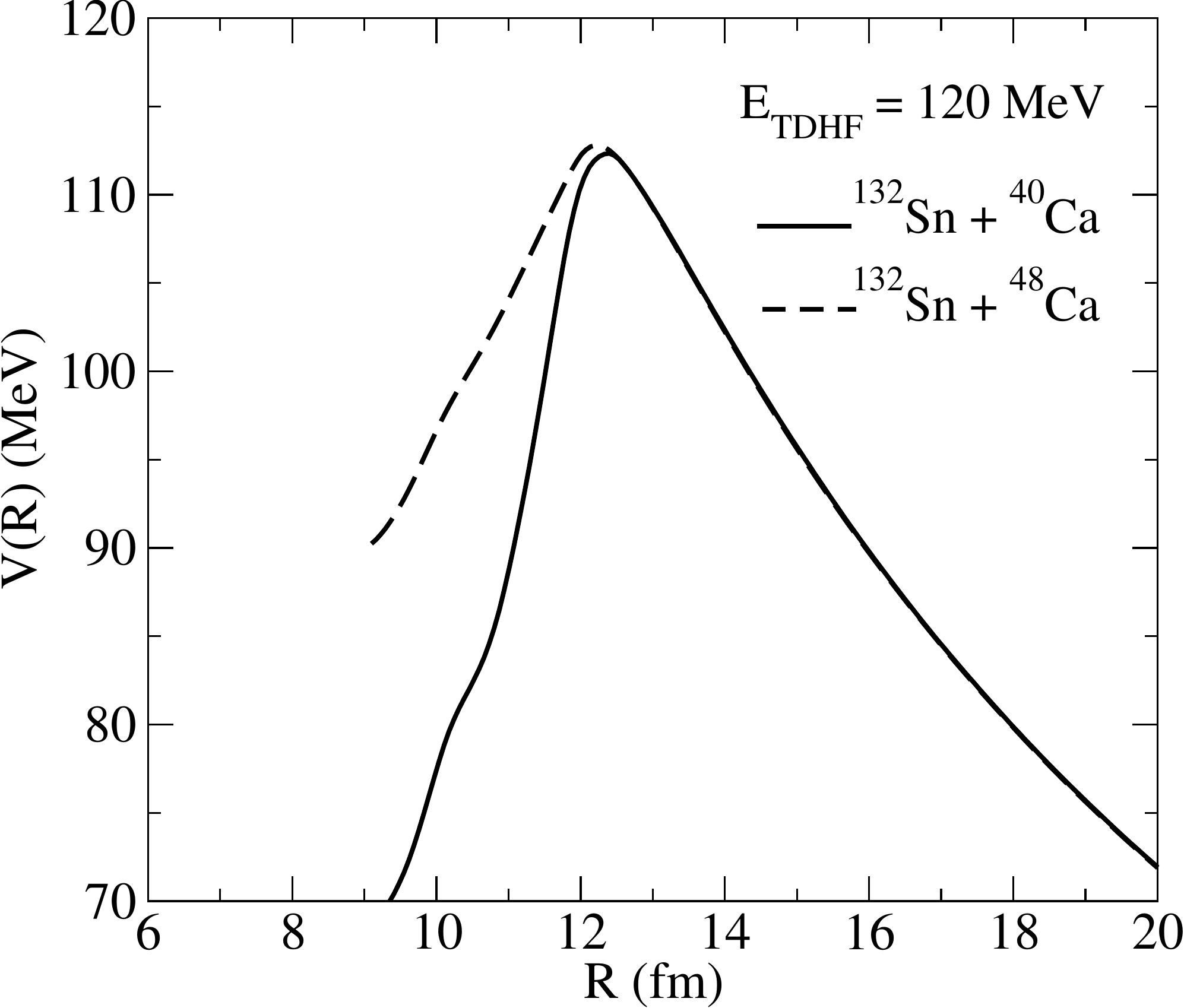}
\caption{DC-TDHF calculation of the heavy-ion potentials $V(R)$ for the systems
$^{132}$Sn+$^{40,48}$Ca. The ion-ion potentials are energy-dependent and have been
calculated at $E_\mathrm{TDHF}=120$ MeV.}
\label{fig1}
\end{figure}

In Fig.~\ref{fig1} we compare the heavy-ion interaction potentials $V(R)$
for the systems $^{132}$Sn+$^{40,48}$Ca. It should be noted that DC-TDHF
contains particle transfer ``on average''~\cite{KU12}, but it does not
describe individual transfer channels. We find the unexpected
result that the barrier heights and positions are approximately the same
in both cases, but the {\it barrier width} for $^{132}$Sn+$^{40}$Ca is
substantially smaller.

Using TDHF dynamics, it is also possible to compute the corresponding coordinate
dependent mass parameter $M(R)$~\cite{UO09b}.
At large distance $R$, the mass $M(R)$ is equal to the
reduced mass $\mu$ of the system. At smaller distances, when the nuclei overlap, the
mass parameter generally increases. We find that the structure of $M(R)$
for the $^{132}$Sn+$^{40}$Ca reaction is fairly similar to the mass
parameter calculated for $^{132}$Sn+$^{48}$Ca (see Fig.~$6$ of Ref.~\cite{OU12}),
and it is therefore not shown here.

Instead of solving the Schr\"odinger equation with coordinate dependent
mass parameter $M(R)$ for the heavy-ion potential $V(R)$, it is numerically
advantageous to use the constant reduced mass $\mu$ and to transfer the
coordinate-dependence of the mass to a scaled
potential $U(\bar{R})$ using a scale transformation
\begin{equation}
(R, M(R), V(R)) \longrightarrow (\bar{R}, \mu, U(\bar{R})) \ .
\label{eq:mrbar}
\end{equation}
Details are given in Ref.~\cite{UO09b}.

\begin{figure}[!htb]
\includegraphics*[width=8.6cm]{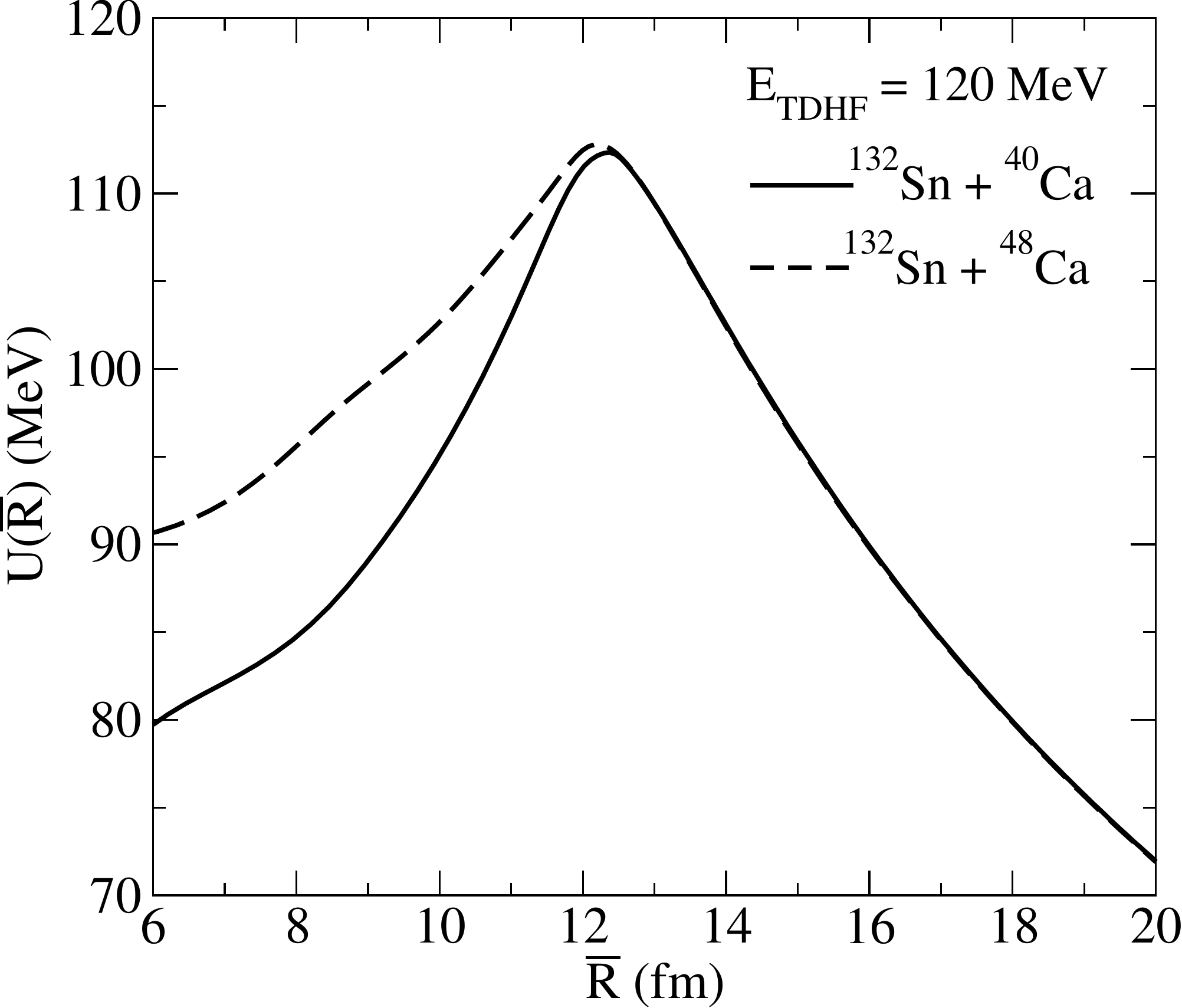}
\caption{Transformed heavy-ion potentials
$U(\bar{R})$ corresponding to the constant reduced mass $\mu$.}
\label{fig2}
\end{figure}

In Fig.~\ref{fig2} we display the transformed potentials $U(\bar{R})$
which correspond to the constant reduced mass $\mu$. A comparison of Fig.~\ref{fig1}
and Fig.~\ref{fig2} reveals that the coordinate-dependent mass changes only
the interior region of the potential barriers, and this change is most pronounced
at low $E_\mathrm{c.m.}$ energies. Note that the transformation to a constant
mass parameter preserves the basic features found for the original potentials,
i.e. the {\it narrower width} of the ion-ion potential for $^{132}$Sn+$^{40}$Ca,
while the barrier heights and positions are approximately the same in both
systems.

\begin{figure}[!htb]
\includegraphics[width=18pc]{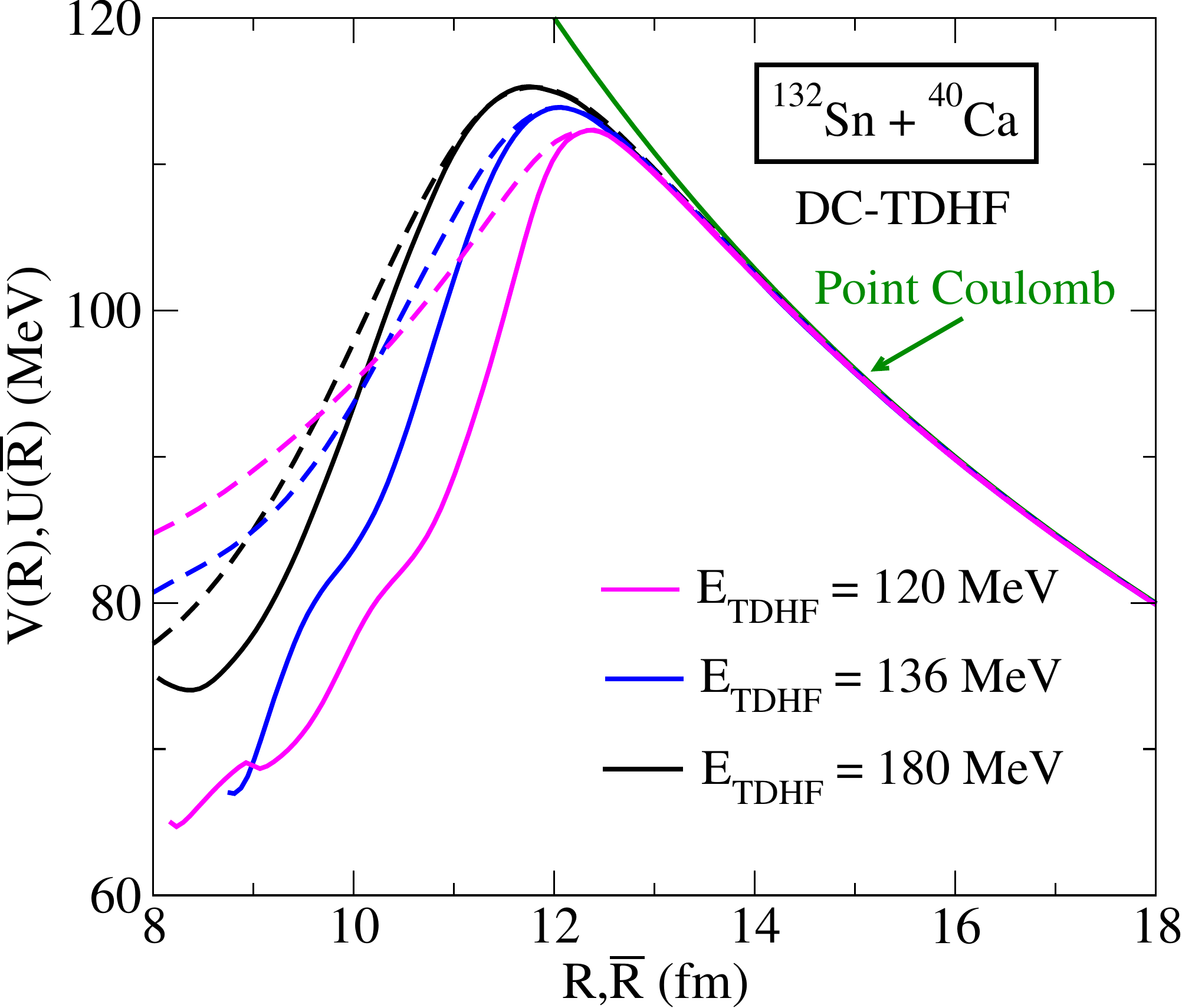}
\caption{(Color online) Solid lines: original heavy-ion potentials $V(R)$. Dashed lines:
transformed potentials $U(\bar{R})$ corresponding to the reduced mass $\mu$.
The potentials have been calculated at three different energies.}
\label{fig3}
\end{figure}

In general, our DC-TDHF calculations show that {\it ion-ion potentials
for heavy systems are strongly energy-dependent}.
By contrast, DC-TDHF calculations for light ion systems such as $^{16}$O+$^{16}$O
show almost no energy-dependence even if we increase $E_\mathrm{c.m.}$ by a factor
of four~\cite{UOMR09}. Even in reactions between a light and a very heavy nucleus
such as $^{16}$O+$^{208}$Pb, we see only a relatively weak energy dependence
of the barrier height and width~\cite{UO09b}.
In Fig.~\ref{fig3} the original potentials $V(R)$ (solid lines) and the
transformed potentials $U(\bar{R})$ are shown at three different TDHF energies.
We notice that in these heavy
systems the potential barrier height increases dramatically with increasing
energy $E_\mathrm{TDHF}$, and the barrier peak moves inward towards
smaller internuclear distances. The potential $U(\bar{R})$ calculated at high energy
($E_\mathrm{TDHF}=180$~MeV) has a barrier $E_B=115.3$ MeV located at $\bar{R}=11.8$ fm,
whereas the potential calculated at low energy ($E_\mathrm{TDHF}=120$~MeV) has a
barrier of only $E_B=112.3$ MeV located at $\bar{R}=12.4$ fm.

The Schr\"odinger equation corresponding to the constant reduced mass $\mu$ and the scaled
potential $U(\bar{R})$ has the familiar form
\begin{equation}
\left [ \frac{-\hbar^2}{2\mu} \frac{d^2}{d\bar{R}^2} + \frac{\hbar^2 \ell (\ell+1)}{2 \mu \bar{R}^2} + U(\bar{R})
 - E_\mathrm{c.m.} \right] \psi_{\ell}(\bar{R}) = 0 \;.
\label{eq:Schroed1}
\end{equation}
By numerical integration of Eq.~(\ref{eq:Schroed1}) using the well-established
{\it Incoming Wave Boundary Condition} (IWBC) method~\cite{HW07} we obtain
the barrier penetrabilities $T_{\ell}$ which determine the total fusion cross
section
\begin{equation}
\sigma_{\mathrm{fus}}(E_{\mathrm{c.m.}}) = \frac{\pi \hbar^2}{2 \mu E_{\mathrm{c.m.}}}
                     \sum_{\ell=0}^{\infty} (2\ell+1) T_{\ell}(E_{\mathrm{c.m.}}) \ .
\label{eq:sigma_fus}
\end{equation}


\section{Fusion cross sections}

\begin{figure}[!htb]
\includegraphics[width=18pc]{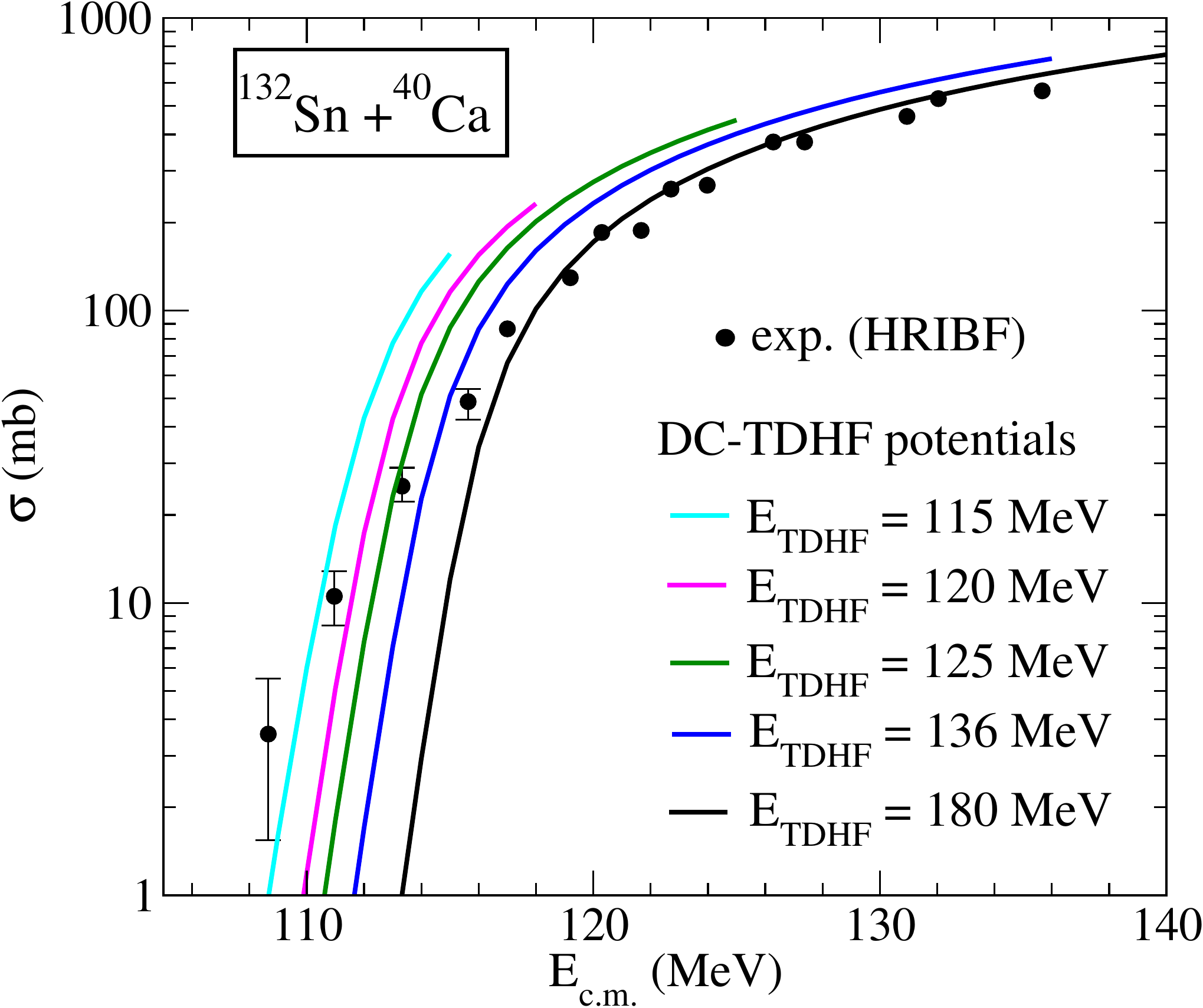}
\caption{(Color online) Total fusion cross sections for $^{132}$Sn+$^{40}$Ca.
The cross sections have been calculated with the DC-TDHF method for several
energy-dependent ion-ion potentials $U(\bar{R})$ some of which are displayed
in Fig.~\ref{fig3}. The experimental data are taken from Ref.~\cite{KR12}.}
\label{fig4}
\end{figure}

\begin{figure}[!htb]
\includegraphics[width=18pc]{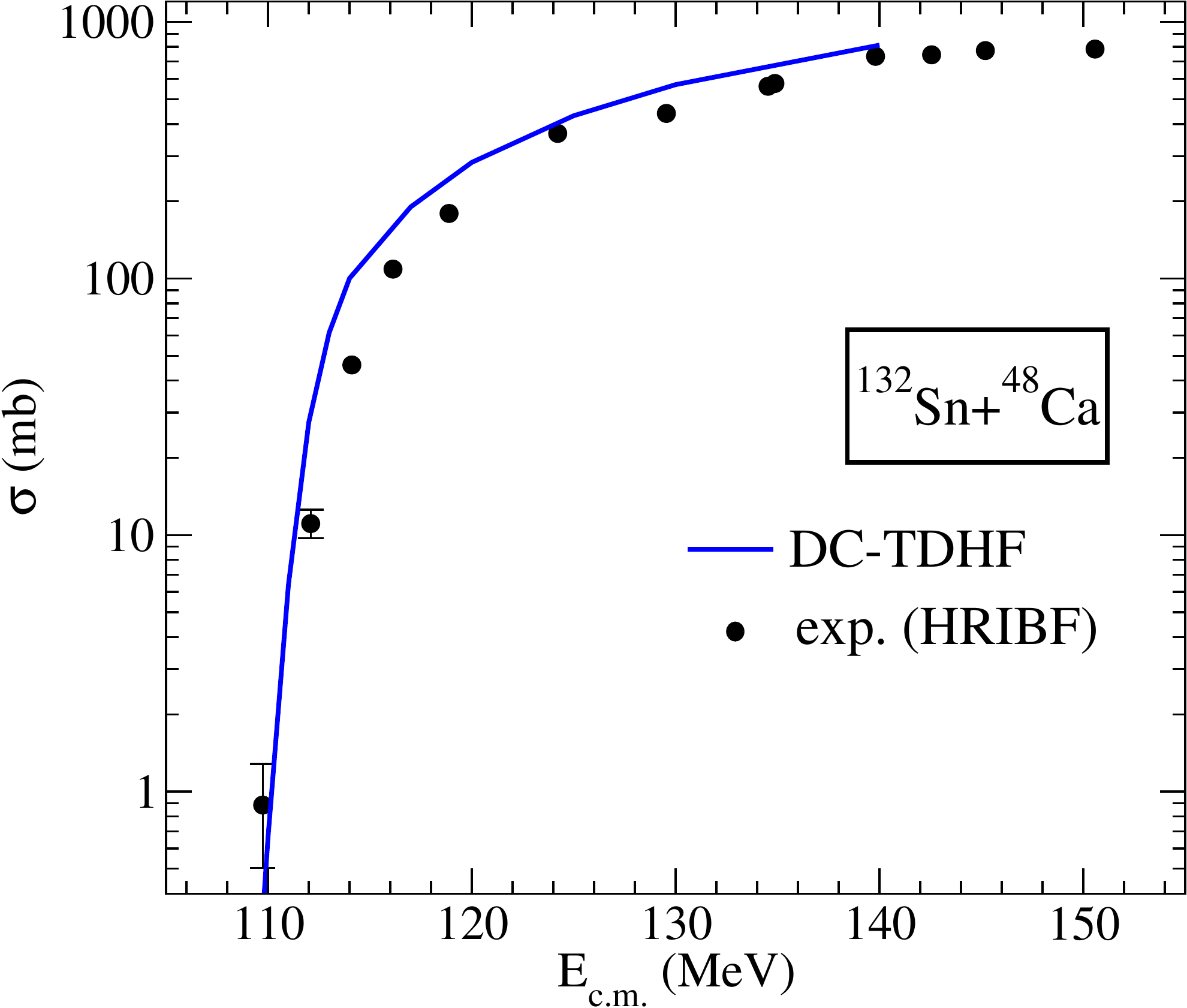}
\caption{(Color online) Total fusion cross section obtained with the
DC-TDHF method for $^{132}$Sn+$^{48}$Ca. Cross sections
calculated for several energy-dependent ion-ion potentials~\cite{OU12} have
been interpolated in this case (single blue line).
The experimental data are taken from Ref.~\cite{KR12}.}
\label{fig5}
\end{figure}

In Figures~\ref{fig4} and~\ref{fig5} we show fusion cross sections measured
at HRIBF~\cite{KR12} for the systems $^{132}$Sn+$^{40,48}$Ca. A comparison of
the fusion cross sections at low energies yields the surprising result
that fusion of $^{132}$Sn with $^{40}$Ca yields a larger cross section than with $^{48}$Ca.
For example, at $E_\mathrm{c.m.}=110$ MeV we find an experimental cross section of $\approx 6$ mb for
$^{132}$Sn+$^{40}$Ca as compared to $0.8$ mb for the more neutron-rich system
$^{132}$Sn+$^{48}$Ca. If the data are scaled for trivial size effects (nuclear radii)
the difference between the ``reduced'' cross sections is found to be even larger,
see Fig.$6$ of Ref.~\cite{KR12}.
The experimentalists have
carried out a coupled channel (CC) analysis of the fusion data with
phenomenological Woods-Saxon potentials which generally underpredict the data
at low $E_{\mathrm{c.m.}}$ energies. In the case of $^{132}$Sn+$^{40}$Ca,
the CC model calculations yield cross sections which differ
by a factor of $10$ or more from the data, despite the fact that
the optical model potentials contain $7$ adjustable parameters.
Interestingly, it is possible to get a good fit to the data using the empirical
Wong model (tunneling through a single parabolic barrier, with $3$
adjustable parameters). In this case the analysis reveals an unusually large
curvature of the barrier, $\hbar \omega = 13.13 \pm 1.09$ MeV, for $^{132}$Sn+$^{40}$Ca
as compared to the $^{132}$Sn+$^{48}$Ca system with only $\hbar \omega = 5.77 \pm 0.52$ MeV.
A large curvature implies a narrow parabolic barrier.
Of course, these values are simply fits to the measured data; the model does
not explain why the barrier curvatures are so dramatically different.
As we will see, our microscopic DC-TDHF theory describes these results naturally
in terms of the underlying mean-field dynamics, without any adjustable
parameters.

In Fig.~\ref{fig4} we show the excitation function of the total fusion cross
section for $^{132}$Sn+$^{40}$Ca. The cross sections have been calculated
with five different energy-dependent ion-ion potentials; the corresponding
DC-TDHF energies are listed in the figure.
The main point of the chosen display is to demonstrate
that the energy-dependence of the heavy-ion potential is crucial for an
understanding of the strong fusion enhancement at subbarrier energies:
At very high energy ($E_\mathrm{TDHF}=180$~MeV) the potential approaches
the limit of the frozen density approximation: the collision is so fast
that the nuclei have no time to rearrange their densities.
We observe that the measured fusion cross sections at energies $E_\mathrm{c.m.}>118$~MeV
are well-described by this high-energy potential.

As we have seen in Fig.~\ref{fig3}, the potential barrier height
decreases dramatically as we lower the energy $E_\mathrm{TDHF}$.
Due to the slow motion of the nuclei at sub-barrier energies, the nuclear densities have time to
rearrange resulting in neck formation, neutron transfer, and
surface vibrations. All of these effects are included in DC-TDHF, and
apparently they reduce the fusion barrier height and strongly modify
the interior region of the heavy-ion interaction potential.
As a result of the decrease in the fusion barrier,
one finds strongly enhanced fusion cross sections at low sub-barrier energies.
In particular, we observe that the data points measured at the two lowest
energies $E_\mathrm{c.m.}=108.6$ and $111$~MeV are described well by the
heavy-ion potential calculated at $E_\mathrm{TDHF}=115$~MeV. This is the
lowest-energy potential $U(\bar{R})$ we have been able to calculate using
the DC-TDHF method, with a
potential barrier $E_B=111.5$ MeV located at  $\bar{R}=12.4$ fm.

In Fig.~\ref{fig5} we show total fusion cross sections for $^{132}$Sn+$^{48}$Ca
which contains $8$ additional neutrons. In this case, we have interpolated the theoretical
cross sections obtained with the energy-dependent DC-TDHF potentials~\cite{OU12}. We can see that our
theoretical cross sections agree remarkably well with the experimental data.
The main experimental puzzle, i.e. the fact that the low-energy sub-barrier fusion cross
section for $^{132}$Sn+$^{40}$Ca is substantially enhanced as compared to
the more neutron-rich system $^{132}$Sn+$^{48}$Ca, can be understood by examining
the transformed ion-ion potential shown in Fig.~\ref{fig2}. Both systems are found
to have approximately the same barrier heights and positions, but the barrier
for $^{40}$Ca has a {\it narrower width}, resulting in enhanced fusion.
Our DC-TDHF approach naturally explains the results of the experimental data
analysis which used the empirical Wong model fit (see remarks at the
beginning of this section). However, the microscopic
potential barrier is not a simple parabola, and the DC-TDHF ion-ion potential
is found to depend strongly on the energy $E_\mathrm{c.m.}$ for heavy systems.


\section{Summary}

In this paper, we calculate heavy-ion interaction potentials and total fusion
cross sections for $^{132}$Sn+$^{40}$Ca at energies $E_{\mathrm{c.m.}}$ below
and above the Coulomb barrier, and we compare these results to the more
neutron-rich system $^{132}$Sn+$^{48}$Ca studied earlier~\cite{OU12}.
The ion-ion potential calculations are carried out utilizing a dynamic microscopic
approach, the Density Constrained Time-Dependent Hartree-Fock (DC-TDHF) method.
The single-particle wave functions are generated on a 3D Cartesian lattice
which spans $50$~fm along the collision axis and $30-42$~fm in
the other two directions. The only input is the Skyrme N-N interaction, there
are no adjustable parameters.

The main objective of this paper is to give a theoretical analysis of fusion cross sections
which were measured recently at HRIBF. The experimental data show a surprising enhancement
at low $E_\mathrm{c.m.}$ energies for the system $^{132}$Sn+$^{40}$Ca as compared to the
more neutron-rich system $^{132}$Sn+$^{48}$Ca. Based on geometric considerations, one
would expect the opposite: as a result of the increased nuclear radius for $^{48}$Ca,
the fusion barrier for $^{132}$Sn+$^{48}$Ca should be reduced which in turn should
increase the fusion cross section. Using the microscopic DC-TDHF approach
we are able to explain the measured sub-barrier fusion enhancement in terms
of the {\it narrower width} of the ion-ion potential for $^{132}$Sn+$^{40}$Ca,
while the barrier heights and positions are approximately the same in both systems.

While for the fusion of light nuclei the microscopic
ion-ion potentials are almost independent of the $c.m.$ energy,
for heavier systems a strong energy-dependence is observed. With increasing
$c.m.$ energy, the height of the potential barrier increases, and
the barrier peak moves inward towards smaller internuclear distances (see Fig.~\ref{fig3}).
This behavior of the ion-ion potential has a dramatic influence on the sub-barrier
fusion cross sections. For the system $^{132}$Sn+$^{40}$Ca, we have calculated
heavy-ion interaction potentials at $8$ different energies ranging from
$E_\mathrm{TDHF}=115$~MeV to $E_\mathrm{TDHF}=180$~MeV.
The time-dependent and density constraint calculations are computationally very
intensive. The total CPU time in this case amounts to $192$ days on a single Intel Xeon processor.
Our calculations are performed on a Dell LINUX workstation with $12$ processors using OpenMP, which
reduces this time to about $16$ days.


\begin{acknowledgments}
This work has been supported by the U.S. Department of Energy under Grant No.
DE-FG02-96ER40975 with Vanderbilt University.
\end{acknowledgments}


\end{document}